\documentclass[12pt]{article}
\usepackage{graphicx}

\begin{document}

\begin{center}
\vspace{1cm}

{\Large\bf International Colloquium}

\vspace{1cm}

{\LARGE\bf Scattering and Scintillation in RadioAstronomy}

\vspace{1cm} {\Large\bf Abstracts}
\end{center}
\vspace{1cm}

{\large International Colloquium  "Scattering and Scintillation in
Radio Astronomy" was held on June 19-23, 2006 in Pushchino, Moscow
region, Russia.

\vspace{1cm} Topics of the Colloquium:}

{\large
\begin{itemize}
\item Interplanetary scintillation
\item Interstellar scintillation
\item Modeling and physical origin of the interplanetary and the interstellar plasma
turbulence
\item Scintillation as a tool for investigation of radio sources
\item Seeing through interplanetary and interstellar turbulent media
\end{itemize}
\vspace{1cm}

Ppt-presentations are available on the Web-site:

http://www.prao.ru/conf/Colloquium/main.html }

\newpage

\begin{center}

{\large\bf
Review of theory of interplanetary and interstellar scintillation
}

{\it
V.I. Shishov
}

Pushcino Radio Astronomy Observatory of P.N. Lebedev Physical Institute, Russia

\end{center}

Basic theoretical methods and solutions described  radio wave propagation in a turbulent plasma are reviewed. Consideration is given to the results on scattering effects such as angle scattering, pulse broadening and spectral line spread. Also it is treated phase and frequency fluctuations. Of particular concern are problem of the correlation theory of intensity fluctuations for weak and strong diffractive and refractive scintillations. Much attention is paid to effect of refractive scintillation on a diffracitive pattern. The distribution function of intensity fluctuation for different regimes of scintillation is discussed.

{\it Acknowledgements}
Authors are grateful for a support of the work by the NSF grant No AST 0098685  and grant No 06-02-16810 of Russian Foundation for Basic Research.

\begin{center}
{\large\bf
Interplanetary Scintillations
}

{\it
W. A. Coles
}

University of California, San Diego, USA
\end{center}

I will discuss the phenomena associated with the scattering of radio waves in the interplanetary plasma, loosely called "interplanetary scintillation" or IPS. This is actually a family of related processes, of which the most important in the solar wind are angular scattering and intensity scintillation. The interplanetary plasma is turbulent and inhomogeneous, so its effect on radio waves must be described statistically. The scattering is caused by fluctuations in electron density wit spatial scales of 10's to 100's of km. These scales are larger than the ion inertial scale, but not much larger, i.e. they lie in the MHD scale range. Angular scattering will cause intensity scintillation if the radio source is sufficiently compact, typically smaller than 1 ur. One can use observations of intensity scintillation to estimate the size of radio sources which are of the order of 1 ur in angular diameter; to study density fluctuations at spatial scales of 10's to 100's of km; to estimate the velocity with which density fluctuations are convected; and to map large scale transient events such as coronal mass ejections as they travel through the solar wind. Observations of angular scattering can be used to study the microstructure in detail, in particular its spatial anisotropy. I will discuss the basic ideas and show some measurements, but I will not discuss measurements in detail since several other speakers will do so later in the program. Finally I will discuss some  more recent ideas resulting from observations near the Sun.

\begin{center}
{\large\bf
Interstellar Scintillation of Pulsars and Extragalactic Radio Sources: a Review
}

{\it
B. J. Rickett
}

University of California, San Diego, USA
\end{center}

Scintillation and scattering are critically important phenomena that must be understood in the study of radio pulsars. As the pulses propagate through the ionized interstellar medium they can be broadened by scattering, their arrival times can wander and their amplitudes scintillate in time and frequency. Pulsar astronomers need to separate these effects from those intrinsic to pulsars. At the same time observations of scintillations allow us to probe the ionized interstellar medium, revealing apparently turbulent structures on scales of 100 km to several AU. I will review progress in both aspects of such scintillation work in the nearly 40 years since pulsars were discovered. In particular, I will describe what can be learned from the extraordinary parabolic arcs that lie hidden in the dynamic spectra of pulsars.

The rich variety of scattering effects are visible in pulsars because of the very small diameters of their emitting regions. While most compact extragalactic sources (quasars, blazars and AGNs) have angular diameters too great to scintillate, some are also small enough to show refractive scintillation over months to years or weak scintillation on time scales from days to hours. I will review extragalactic ISS emphasizing recent observations of the fastest variations and what they tell us about the sources and the medium.

\begin{center}
{\large\bf
Coronal Faraday rotation of occulted radio signals
}

{\it
M. K. Bird
}

AIfA, University of Bonn, Germany
\end{center}

Faraday rotation (FR) observations of radio sources near solar conjunction yield information on the coronal magnetic field at heliospheric distances not reached by in situ expolaration. Measurements of FR yield the rotation measure (RM), a wavelength-independent quantity defined as the integral along the raypath of the product of the electron density times the raypath-parallel component of the magnetic field. Independent observations or models of the coronal electron density are required in order to extract information about the magnetic field. The radio sounding sources can be either artificial (spacecraft) or natural, but they must be at least partially linearly polarized. The most extensive campaign of coronal radio sounding polarization measurements using a spacecraft was the Helios Faraday Rotation Experiment, which was conducted over the duration of the Helios 1 (1974-84) and Helios 2 (1976-80) missions. Other coronal FR experiments have been carried out using natural continuum sources rather recently at the VLA and as early as 1962 at the Pulkovo Radio Telescope. Pulsars were exploited to determine coronal RM at the MPIfR Effeleberg 100-m Telescope. Extending these single-raypath observations, an ambitious project to use the future LOFAR facility for constructing an \'image\' of coronal RM has been proposed. Different time scales of FR variations are related to different physical phenomena. Among the observed effects are: (a) slow variations associated with the changing geometry and rotation of the corona; (b) random oscillations probably arising from a rich spectrum of coronal Alfven waves; (c) rapid changes in RM caused by transient events such as coronal mass ejections (CMEs).

\begin{center}
{\large\bf
Observation of the January 1997 coronal mass ejection near the Sun using radio sounding technique with the GALILEO spacecraft
}

{\it
A.I. Efimov$^1$, L.N. Samoznaev$^1$, V.K. Rudash$^1$, I.V. Chashei$^2$, M.K. Bird$^3$, D. Plettemeier$^4$
}

$^1$ Institute of Radio Engineering and Electronics RAS, Russia

$^2$ Pushchino Radio Astronomy Observatory of the P.N. Lebedev Physical Institute, Russia

$^3$ Argelanges Institute of Astronomy Bonn University, Germany

$^4$ Technical University, Dresden, Germany
\end{center}

Frequency and amplitude fluctuations of the GALILEO S-band radio signal were recorded in January 1997. The strong enhancement of the radio wave fluctuations was detected from 16:20 UT on 8 January to 08:30 UT on 9 January when the radio ray path proximate point was on the west limb at about 32 solar radii from the Sun. The passage of the perturbed plasma flows through the line of sight is characterized by a significant increase of the intensity and frequency fluctuations, a change of the turbulence regime, by two - velocity structure of the plasma flows with the high velocity at about 400 km/s and  the low velocity at about 200 km/s. It is shown that these radio effects were connected with the coronal mass ejection which was observed first in the field of view of the SOHO/LASCO coronagraphs near the Sun on January 6 and were associated with many phenomena observed near the Earth on January 10 - 11, 1997 (jumps of the ion density, peaks of the velocity, enhancements of the interplanetary magnetic field, variations of the global ionospheric TEC etc).

\begin{center}
{\large\bf
Coronal Faraday Rotation: Diagnostics of Current Sheets and MHD Waves
}

{\it
Steven R. Spangler
}

University of Iowa, USA
\end{center}

Interplanetary scintillations yield information on plasma density fluctuations. The present theoretical understanding of magnetohydrodynamic (MHD) turbulence states that these fluctuations are second order responses to the dominant  magnetic field and velocity fluctuations. These latter fluctuations also contain the bulk of the turbulent energy density which can heat the plasma through dissipation. Information on magnetic field and velocity fluctuations is therefore important for understanding plasma processes in the corona, as well as the interplanetary medium and (by extension) the interstellar medium. Measurements of Faraday rotation of linearly polarized spacecraft transmitters and extragalactic radio sources provide information on the magnetic field in the corona and inner solar wind. This Faraday rotation can reach values as high as of order 100 radians per meter squared at a heliocentric distance of 5 solar radii. In addition to information on the large scale, global coronal magnetic field, spatial and temporal variations in the coronal rotation measure give information on plasma inhomogeneity and turbulence. I will discuss what can be learned about inhomogeneity and turbulence from (1) measurements of temporal fluctuations in rotation measure (primarily from Faraday rotation of spacecraft beacons), (2) limits to the depolarizing characteristics of the corona, and (3) differences in the Faraday rotation along two closely spaced lines of sight through the corona.

\begin{center}
{\large\bf
SURA-WIND experiments: intensity fluctuations due to Faraday effect
}

{\it
Yu. Tokarev, Yu. Belov, G. Boiko, G. Komrakov
}

Radiophysical Research Institute, N.Novgorod, Russia
\end{center}

Investigations of near-Earth space environment by observation of
the SURA facility emission onboard the NASA WIND spacecraft are
conducting successfully since August 1995. The SURA signals
registered by WIND RAD2 receiver strongly fluctuate as a rule,
because of radio waves are scattering by the near-earth plasma
irregularities. Low frequency part of fluctuation spectrum
(fluctuation frequencies $f < 0.1$ ~Hz) is produced by ionospheric
scintillations. A high frequency peculiarity of the spectrum
around $f = 1 - 4$ ~Hz could be referred to radio waves scattering
at the solar wind clouds with scale about 40 - 100 km. A special
technique of the SURA-WIND data processing which permits to
separate the ionospheric and interplanetary components of the
scintillation spectrum was developed.

It was detected recently that the remarkable intensity variations
of received signal can be too produced by the Faraday wave
propagation effect of the SURA emission, having a considerable
linear polarization, in the nonstationary Earth ionosphere.
Analysis of the phenomena is complicated because expected fringes
due to polarization plane rotations are usually superimposed on
the strong ionospheric scintillations.

A method, which allows refining the Faraday fluctuations of SURA
signal, using special combination of the RAD2 receiver outputs
connected to linear dipoles, is presented in the report. The
method possibilities are illustrated for some SURA-WIND sessions
with pronounced subsidiary maximum of the fluctuation spectrum
caused by transient conditions in the Earth ionosphere.
Recommendations in order to use the Faraday fluctuations for
diagnostic of magnetoactive space plasma are discussed too.

{\it Acknowledgements}
Authors are grateful for a support of the work by the INTAS grant N 03-51-5727 and grant N 112/001/804 for Support of Leading Scientific Schools.

\begin{center}
{\large\bf
Solar Radar Observations at 50 MHz,
}

{\it
W. A. Coles, J. Chau, J. Harmon, M. Sulzer
}

University of California, San Diego, USA
\end{center}

The Sun was one of the first target's to be attempted when radar technology advanced to the point where extra-terrestrial targets could be contemplated. Observations were made between 1960 and 1969 using a special purpose radar operating at 38 MHz, but these observations were not understood. The solar community lost interest in the technique and the radar was scrapped. We recently tried to resurrect the idea using the large incoherent scatter radar at Jicamarca in Peru. The sensitivity of this system was similar to the original, but we were unable to detect any echoes. We believe that any echoes are much weaker than had been thought, and we suggest several reasons why they might be much weaker.

\begin{center}
{\large\bf
IPS tomographic observations of 3D solar wind structure
}

{\it
M. Kojima, M. Tokumaru, K. Fujiki
}

STE Lab., Nagoya University, Japan
\end{center}

The interplanetary scintillation (IPS) technique is one of the few that can be used to observe the solar wind in three-dimensional space. IPS has several advantages over in situ spacecraft measurements. It can be used consistently for a long-term study of solar wind structure over the solar cycle. In addition, when a large number of IPS sources are available, vast regions of interplanetary space can be probed in a relatively short time. However, since IPS measurements of solar wind properties integrate over the line of sight, the solar wind structures studied using the IPS method were sketchy. We developed tomographic analysis method for IPS observation in order to deconvolve the line-of-sight integration effect. This technique can retrieve not only three dimensional solar wind parameters but also provide high spatial resolution. Nowadays IPS measurements are greatly improved qualitatively from earlier ones. We introduce the technique of the IPS tomographic analysis and the origins of the solar wind revealed by this technique.

\begin{center}
{\large\bf
Low-Density Solar Wind Anomalies
}

{\it
P. Janardhan
}

Physical Research Laboratory, Ahmedabad, India
\end{center}

The solar wind at 1 AU is known to be strongly supersonic and super Alfvenic with average solar wind densities being $\sim 10$ ~cm. However, there have been several occasions when the solar wind densities at 1 AU have dropped to values $< 0.2 cm^{-3}$ for periods ranging from several hours to more than 24 hours. The most dramatic of these events occurred on May 11 and 12 1999, when the Earth was engulfed by an unusually low density ($< 0.1$ ~cm) and low velocity ($< 350$ km s) solar wind with and Alfven Mach number significantly less than 1. This was a unique low-velocity, low-density, sub-Alfvenic solar wind flow, which spacecraft observations have shown lasted more than 24 hours. In this paper we will present IPS observations of some of these events and discuss the possible solar sources for such unusual solar wind flows.

\begin{center}
{\large\bf
Comparison of the Extent and Mass of CME Events in the Interplanetary Medium Using IPS and SMEI Thomson Scattering Observations
}

{\it
B. V. Jackson$^1$, P. P. Hick$^1$, A. Buffington$^1$, M. Kojima$^2$, M. Tokumaru$^2$
}

$^1$ CASS/UCSD, USA

$^2$ STE Lab., Nagoya University, Japan
\end{center}

The Solar-Terrestrial Environment Laboratory (STELab), Japan interplanetary scintillation (IPS) g-level and velocity data measurements give the extent of CME disturbances in the interplanetary medium from intensity scattering from the intervening medium. White-light Thomson scattering observations from the Solar Mass Ejection Imager (SMEI) have recorded the inner heliospheric response to several hundred CMEs. In this presentation we detail the differences between these two techniques by measurement of the extent of several well-observed CMEs during the SMEI data period (February 2003 - present) using Thomson scattering from electrons at the same elongations as IPS g-level observations, and show how these measurements compare in sky maps. For SMEI data we employ a 3D reconstruction technique that obtains perspective views from outward-flowing solar wind as observed from Earth, iteratively fitting a kinematic solar wind density model using the SMEI white light observations and, when available IPS velocity data.  This technique helps separate the heliospheric response in SMEI from other sources of background noise, and also provides the 3D structure of the CME and its mass.  For instance, the analysis shows and tracks outward the northward portion of the loop structure of the October 28, 2003 halo CME observed in LASCO images that passes Earth on October 29. We determine an excess 3D mass for this structure of $6.7 \times 10^{16}$ ~g and a total mass of $8.3 \times 10^{16}$ ~g in the SMEI data, and this compares with similar values obtained by M. Tokumaru of STELab using IPS scintillation-level data and a 3D reconstruction technique developed for these data and applied to this event.  We further explore this relationship in our analysis, and in addition, show further application for these analyses.

\begin{center}
{\large\bf
Global properties of heliospheric disturbances observed by interplanetary scintillation
}

{\it
M. Tokumaru
}

STE Lab., Nagoya University, Japan
\end{center}

Daily observations of interplanetary scintillation (IPS) using celestial sources suitably distributed over the sun-centered sky plane allow us to image the solar wind plasma between the corona and the earth orbit. This IPS imaging technique is particularly useful to study physical properties of the large-scale heliospheric disturbance. We analyzed IPS data collected with the 327-MHz four-station system of the Solar-Terrestrial Environment Laboratory of the Nagoya University in order to gain insight into the 3-dimensional structure and propagation dynamics of solar wind transients associated with prominent coronal mass ejection (CME) events. As result, global features with an elongated distribution were deduced for some CME events, while nearly isotropic features were observed for the other events.

The variety of global features may be ascribed to either ambient solar wind conditions or properties of the coronal ejecta. In addition, propagation speeds of the transients were found to evolve significantly during the propagation, and the radial evolution appeared to depend on the ambient solar wind conditions.

\begin{center}
{\large\bf
To the possibility of research of the external solar wind thin structure in decameter radio waves
}

{\it
M. R. Olyak
}

Institute of Radio Astronomy, Kharkov, Ukraine
\end{center}

The Feynman path-integral technique was applied for calculations of the cross - spectra of weak interplanetary scintillations. The possibility of study of the thin structure of the solar wind external areas by the simultaneous measurements of temporary scintillation spectra and the phase speed dispersion dependencies is shown for the decameter range of waves.

\begin{center}
{\large\bf
IPS Using EISCAT and MERLIN: Extremely Long Baseline Observations at Multiple Frequencies
}

{\it
R. A. Fallows
}

University of Wales, Aberystwyth, UK
\end{center}

An upgrade to two of the radio telescopes of EISCAT in northern Scandinavia to enable measurements of IPS to be taken at 1.4GHz has prompted two major developments in studies of IPS by the Aberystwyth Solar System Physics group in recent years. Simultaneous observations between EISCAT and MERLIN in the UK allow for baselines of up to 2000km showing a further improved velocity resolution, much greater accuracy in determining the direction of flow than previously, and demonstrating that density variations in the slow solar wind remain correlated for at least 8s. Initial results indicate two fast solar wind modes and suggest that the direction of flow changes according to the solar wind configuration in the line of sight, with deviation from radial of up to 4 degrees seen. Correlating measurements at different observing frequencies has been trialled, including the use of the EISCAT Svalbard Radar (ESR) for IPS for the first time. Observations at 500MHz, 928MHz and 1.4GHz with baselines of up to 1200km have been carried out. The results are found to be consistent between single- and dual- frequency correlations, allowing the range of observations possible with the EISCAT system to be expanded.

\begin{center}
{\large\bf
Theoretical and experimental investigations of solar wind plasma using of VLBI-method
}

{\it M.B. Nechaeva, V.G. Gavrilenko, Yu.N. Gorshenkov, V.A.Alimov,
I.E. Molotov, A.B. Pushkarev, R. Shanks, G.Tuccari}

Radiophysical Research Institute, Russia
\end{center}

We report the results of theoretical and experimental works on investigation of solar wind plasma by the method of radio raying with using of Very Long Baseline Interferometer (VLBI). The results of two VLBI-experiments on wavelength 18 cm are discussed. Observations were performed on the basis of LFVN (Low Frequency VLBI Network) in 1999 and 2000 with using S2 registration system and was processed on S2-correlator at Penticton (Canada). The post processing, carried out in Radiophysical research institute (Nizhnij Novgorod, Russia) was aimed to obtain value of solar wind velocity and index of spatial spectrum of electron density fluctuations. Experimental results were compared with theoretical conclusions.

\begin{center}
{\large\bf
Structure of solar wind flows at the maximum and descending phases of the 23 solar activity cycle
}

{\it
N. A. Lotova$^1$, K. V. Vladimirski$^2$, V.N. Obridko$^1$
}

$^1$ IZMIRAN, Russia

$^2$ Lebedev Physical Institute, Russia
\end{center}

The structure of solar wind streams is investigated on the base of combined analysis of data on location of inner boundary of the solar wind transsonic transition region Rin, calculations of the coronal magnetic field structure and strength, and white corona images obtained by the SOHO LASCOC2. Analysis of the observational data on the streams structure for the 2000-2002 solar maximum period and 2003-2004 descending period has shown that the transition year from maximum to descending phase does not coinside with it's definition using Wolf numbers Rz as well as using intencity of the green coronal line.

\begin{center}
{\large\bf The identification of the fluctuation effects related
to the turbulence and ''permanent'' layers in the atmosphere of
Venus from radio occultation data. }

{\it
V.N. Gubenko, V.E. Andreev
}

Institute of Radio Engineering and Electronics, Russian Academy of Sciences, Russia
\end{center}

The results of cross-correlation analysis of the amplitude
fluctuations of radio waves of the $\lambda = 32$ ~cm band in
seven sessions of radio occultation measurements of the northern
polar atmosphere of the planet are presented. The existence of the
cross-correlation of fluctuations ($b_{\xi} \cong 0.6-0.7$) is
established in the altitude realizations in the interval 61.5-65.0
km for 4 different sessions of radio occultation. Inner layering
is revealed in the upper layer of the clouds of the planet at
altitudes of 61.5-65.0 km, which is specified by an enhanced
turbulence of the atmosphere. It is found that the ''lifetime'' of
the small-scale layered irregularities is no less than 2 days and
that their horizontal extension in the meridional direction can
exceed $\sim 130$ ~km. A possible cause of the emergence of the
layered structures inside the upper layer of the polar clouds of
Venus is discussed.

\begin{center}
{\large\bf
Using pulsar scintillation to probe AU-size structure in the interstellar medium
}

{\it
D. Stinebring
}

Oberlin College, USA
\end{center}

It is well known that pulsar dynamic spectra occasionally show pronounced fringing or crisscross patterns. It was a surprise, however, that a 2d Fourier analysis of these spectra showed faint, parabolic features in the secondary spectra, which are now called scintillation arcs. I will show evidence that the scintillation arc phenomenon is widespread and that it underpins many other scintillation phenomena. If an estimate of the distance to the pulsar and a measurement of its proper motion exist, the location of the scattering material along the line of sight can be determined. There is often pronounced substructure in the arcs, and it translates along the main arc in a manner that is determined by the proper motion of the pulsar. This substructure may be produced by lens-like features in the ionized ISM that are far out of pressure balance with the warm ionized medium (WIM) and that may be related to deterministic structures that cause extreme scattering events.

Observations with this technique, which rely on a large flux density and/or a large collecting area, have an angular resolution of about a milliarcsecond. They often show features in the scatter broadened image out to 15 -- 20 times this resolution, however. Thus, single-dish observations can study details in the scattering medium on AU-size scales while covering a relatively large field of view that scans the sky at the pulsar proper motion speed. We are still learning how to interpret the richly detailed scintillation arc pattern that results, and bservational and interpretive surprises continue to emerge.

\begin{center}
{\large\bf
Interstellar Levy Flights
}

{\it
C. Gwinn
}

University of California, Santa Barbara, USA
\end{center}

We present a model for interstellar scattering via non-Gaussian statistics, resulting in Levy flight, and compare it with observations. In the Levy model, rare, large events dominate statistical averages. Consequently, the distribution of wave directions, after many scatterings, approaches a Levy distribution. Like Gaussian distributions, such distributions are stable under repeated convolution, and indeed are attractors; however, they do not have finite moments. For example, physical models where density differences between points are drawn from distributions without finite second moments lead to Levy statistics for scattering. Such models may include the scaling of density differences with separation of points that is consistent with Kolmogorov statistics.

Levy flights may help to rephrase, or resolve, some long-standing paradoxes of interstellar scattering. They can explain the scaling of pulsars' pulse broadening tau with dispersion measure DM as $tau \sim DM^4$, using stationary statistics (the "Sutton paradox"). They can explain the sharp rise time and long tail of scatter-broadened pulses, with a uniform medium ("Williamson paradox"). They can explain the sharp cusps and extended halos, relative to the traditional Gaussian Kolmogorov model, of scatter-broadened images ("Desai paradox"). We present an overview of theory and comparisons with recent observational tests.

\begin{center}
{\large\bf
The MASIV VLA 5 GHz Scintillation Survey of the Northern Radio Sky
}

{\it
J.E.J. Lovell, D.L. Jauncey, C. Senkbeil, S. Shabala, H.E. Bignall, J-P. Macquart, B.J. Rickett, L. Kedziora-Chudczer
}

ATNF, Australia
\end{center}

We are analyzing the results of a large-scale, 550 flat-spectrum sources,
5 GHz VLA Survey of the northern sky searching for sources that exhibited intra-day variability (IDV). The observations were taken over four epochs each of the three days duration in January, May, September 2002 and January 2003 during VLA reconfiguration time. The objective was to obtain a large sample, $\sim$ 100 or more sources, that exhibited IDV in order to derive reliable statistics on the microarcsecond properties of the source population and the scattering properties of the interstellar medium. The survey was named the Micro-Arcsecond Scintillation-Induced Variability survey, or MASIV. We present the results of our analysis.

\begin{center}
{\large\bf
Interstellar turbulent plasma spectrum from multi-frequency
observations of pulsars
}

{\it
T. V. Smirnova
}

Pushchino Radio Astronomy Observatory of P.N. Lebedev Physical Institute, Russia
\end{center}

We will report about our recent results concerning the shape of interstellar plasma spectrum based on complex analysis of multi-frequency observations of diffractive and refractive scintillation of pulsars. Although the Kolmogorov spectrum describes data sufficiently well in a statistical sense, the dispersion of points is large and we found that in the particular directions the spectrum differs from the Kolmogorov one.

\begin{center}
{\large\bf
Interstellar Scintillation of the Double Pulsar J0737-3039
}

{\it
B. J. Rickett, W. A. Coles
}

University of California, San Diego, USA
\end{center}

The double pulsar is not only a remarkable laboratory for general relativity, late stage evolution of binary stars and pulsar magnetospheres, it is also a unique laboratory for interstellar scintillation (ISS). We have observed the ISS of both pulsars over one year at 1.7-2.2 GHz using the Green Bank Telescope. From these data (and other data from the literature) we are studying the orbital modulation of the ISS timescale, the correlation in the ISS signals from the two pulsars and the effects of large scale gradients in the electron column density.
We present the theory of these effects and show that the observations require that the interstellar turbulence be anisotropic and the center of mass velocity be lower than had first been thought. The inclination of the pulsars' orbit is so close to 90 deg that orbital modulation of ISS time scale degenerates to a form characterized by only three parameters. However, the timescaale depends on five physical quantities: the center of mass velocity of the pulsars, the spatial scale, axial ratio and orientation of the scattering plasma. In cases where we can observe the correlation between the ISS of the A and B pulsars we gain extra observables but not enough to determine all five physical parameters. Although we cannot fit all the unknowns at each epoch, we can use the changes in the Earth's orbital velocity over a year to provide extra degrees of freedom. In this way, we can constrain the true center of mass velocity in a known reference frame; the axial ratio and orientation of the scintillation; and the distance to the interstellar scattering layer. We will present the latest conclusions from this project.

\begin{center}
{\large\bf
Diffraction scintillation at 1.4 and 4.85 GHz
}

{\it
V. M. Malofeev$^1$, O. I. Malov$^1$, S. A. Tyulbashev$^1$, A. Jessner$^2$, W. Sieber$^3$, R. Wielebinski$^2$
}

$^1$ Pushchino Radio Astronomy Observatory of P.N. Lebedev Physical Institute, Russia

$^2$ Max-Planck-Institut f\" ur Radioasrronomie, Bonn, Germany

$^3$ Hochschule Niderrhein, Krefeld, Germany
\end{center}

Investigation of the intensity fluctuations, caused by interstellar scintillations, for 18 pulsars at 1.4 and 4.85 GHz is presented. Observations have been made at the 100-m radio telescope of the Max-Planck-Institut fur Radioastronomie. Most of observations have carried out during a few hours and using both the broad 80 MHz and 200-500 MHz bandwidth at 1.4 and 4.85 GHz correspondingly and also the 30 $\times$ 1.33 MHz filter bank at 1.4 GHz and the 8 $\times$ 60 MHz ones at 4.85 GHz. It gives the possibility to measure not only temporal scintillation parameters: scintillation indices and decorrelation times, but also the decorrelation bandwidth and dynamic spectra at so high radio frequencies.

\begin{center}
{\large\bf
The Dedicated Interferometer for Rapid Variability (DIRV)
}

{\it
B. K. Dennison, C. A. Bennett, R. M. Blake, M. W. Castelaz, D. Cline, C. S. Osborne, L. Owen, W. A. Christiansen, D. A. Moffett
}

University of North Carolina, Asheville, USA
\end{center}

A project is underway to develop a two-element interferometer for dedicated monitoring of compact radio sources. The primary goal will involve long-term, fully-sampled monitoring of a large sample of extragalactic sources to study interstellar scintillation and search for extreme scattering events. The interferometer will make use of two existing 26-meter antennas with an east-west separation of about 400 meters at the Pisgah Astronomical Research Institute (PARI), in partnership with the Pisgah Astronomical Research and Science Education Center (PARSEC), a center of the University of North Carolina system. This research is supported by PARI, the US National Science Foundation through grant AST-0520928 to UNCA, and the Glaxo-Wellcome Endowment at UNCA.

\begin{center}
{\large\bf
Interstellar scintillation as a probe of muas structure in quasars
}

{\it
H. E. Bignall
}

Joint Institute for VLBI in Europe, Netherlands
\end{center}

Observations over the last two decades have shown that a significant fraction of all flat-spectrum, extragalactic radio sources exhibit flux density variations on timescales of a day or less at frequencies of several GHz. It has been demonstrated that interstellar scintillation (ISS) is the principal cause of such rapid variability. Observations of ISS can be used to probe very compact, microarcsecond-scale structure in the AGN, as well as properties of turbulence in the local Galactic ISM. A few quasars have been found to show unusually rapid, intra-hour variations, evidently due to scattering in very nearby ($\sim$ 10pc), localized turbulent plasma. For these sources, it is relatively easy to study the ISS in detail as the scintillation pattern is well sampled in a typical observing session. The recent large-scale MASIV VLA Survey showed that such rapid ISS is extremely rare, and thus monitoring over much longer periods is required to study the ISS of most quasars in similar detail. Some recent observational results are reviewed, and methods and problems for using ISS as a probe of quasar structure are discussed.

\begin{center}
{\large\bf
Probing Cosmic Plasma with Giant Radio Pulses,
}

{\it
V.I. Kondtratiev, M.V. Popov, V.A. Soglasnov, N. Bartel, W. Cannon, A.Y. Novikov
}

Astro Space Center of P.N. Lebedev Physical Institute, Russia
\end{center}

VLBI observations of the Crab pulsar were conducted at 2.4 GHz using 64-m Kalyazin radiotelescope and 45-m dish at Algonquin (Canada) with S2 recording system. The data were processed with software correlator, which permits to obtaine visibility on single giant pulses. Behavior of visibility delay and phase with time gave an unique information on the turbulent plasma near the pulsar, inside the SNR nebula, and in ISM. New unexpected propagation effects are revealed, in particular, a different scattering for RCP and LCP modes.

\begin{center}
{\large\bf
Measurements of the scatter broadening of pulsar radio emission and a homogeneity of the turbulent plasma in the near Galaxy
}

{\it
A. D. Kuzmin, B. Ya. Losovsky
}

Pushchino Radio Astronomy Observatory, Lebedev Physical Institute, Russia
\end{center}

Pulsars as the point sources of a pulsed radio emission, distributed over the whole Galaxy, are a good instrument for investigation the scatter and a turbulent plasma properties in our Galaxy.
We report the low frequency measurements of the scatter pulse broadening  of a large sample of pulsars in a vast Galaxy region of the galactic longitude from $6^{\circ}$ to $252^{\circ}$ and the distance up to 3 kpc.

Low frequency measurements  (110, 60 and 40 MHz) expand several times a frequency interval  of  $\tau_{sc}$ data and provide precise determination of the frequency dependence $\tau_{sc} (v) \propto v^{4.0 \pm 0.3}$.
Large sample of 100 pulsars and uniform measurements and reduction processes
provide  precise determination of a dispersion measure dependence $\tau_{sc} (DM) \propto DM^{2.2 \pm 0.3}$.

The vast scope of the Galaxy provide a comparison of a turbulence level Cn$^2$ in  various directions and distances and have revealed that  the Galaxy turbulent plasma in this region is rough homogeneous one.

\begin{center}
{\large\bf
Distribution of the turbulent plasma in the Galaxy
}

{\it
A. V. Pynzar
}

Pushchino Radio Astronomy Observatory, Lebedev Physical Institute, Russia
\end{center}

Distribution in the Galaxy of the parameter  $\tau / (DM)^2$   (where  $\tau$  is the pulsar pulse broadening due to scattering in the interstellar medium and DM - dispersion measure)  is investigated. $\tau / (DM)^2$   characterizes the relative level of the electron  density fluctuations  in the interstellar turbulent plasma. It is confirmed, that  $\tau/(DM)^2 \propto (DM)^4$. This fact shows, that  parameters $\tau$ and DM are due to thesame regions of the interstellar medium. It is shown,that  $\tau/(DM)^2$   increasesstrong as the galactic latitude and longitude decreases. The characteristic  distribution scales are $1^{\circ}$ in latitude and $30^{\circ}$ in longitude. It is revealed, that $\tau/(DM)^2$   increases significantly as angular distance between pulsars  and supernova remnants decreases less $1^{\circ}$. It is drawn a conclusion, that turbulence is strong near HII regions surrounded supernova remnants. Turbulenc forming mechanisms and energy origins are discussed.

\begin{center}
{\large\bf
The distribution of magnetic field in the plane of galaxy
}

{\it
R. R. Andreasyasian
}

Byurakan Astrophysical Observatiry, Armenia
\end{center}

It was studied the distribution of magnetic field in the plane of Galaxy by using the data of Faraday rotation measure (RM) of low latitude pulsars. We have constructed the detail maps of magnetic field in coordinates (l;R) and also in (l;DM), where l - is galactic longitude, R- is the distance from the Sun and DM- is dispersion measure. For the construction of maps we used a new procedure of averaging, and calculating of the magnetic field value in every point of Galactic plane. From the maps we find the correlations of magnetic field with the known spiral arm structure of Galaxy and determine the accurate positions of magnetic field reversals. It must be noted, that for the construction of detail maps of magnetic field in coordinates (l;DM) we used only the coordinates of pulsars and their primary observational data (RM and DM).

\begin{center}
{\large\bf
A Galactic Latitude - Modulation Index relation as an indicator of the ISM scintillation of the extragalactic sources radio emission
}

{\it
G. S. Tsarevsky
}

Astro Space Center of P.N. Lebedev Physical Institute, Russia
\end{center}

We show that massive variability surveys of the extragalactic radio sources taken at different frequencies display a prominent trend in the Galactic Latitude - Modulation Index ($|b| - m$) plot. Taken alone, it has a pretty small statistical significance, but as a pool proves directly the extrinsic, ISM scintillation origin of the variability, at least partly. To discriminate each other - the intrinsic and extrinsic parts - is a challenge for the scintillation theory.

\begin{center}
{\large\bf
Solar wind turbulence from radio occultation data
}

{\it
I. V. Chashei$^1$, A. I. Efimov$^2$, M. K. Bird$^3$
}

$^1$ Pushchino Radio Astronomy Observatory of the P.N. Lebedev Physical Institute, Russia

$^2$ Institute of Radio Engineering and Electronics RAS, Russia

$^3$ Argelanges Institute of Astronomy,  Bonn University, Germany
\end{center}

The characteristics of plasma turbulence in the inner solar wind, as deduced from radio frequency fluctuation measurements recorded during solar conjunctions, are reviewed. Special emphasis is placed on the results from radio occultation experiments performed with the Galileo and Ulysses spacecraft in the interval 1991-2002. Estimates of the power spectral index and of the turbulence outer scale are obtained in the range of heliocentric distances $5 R_{|} < R < 80 R_{|}$. The radial evolution of these parameters is discussed. The turbulence properties in the low latitudes slow solar winds are shown to be invariant over the solar cycle. The observations are interpreted within the framework of a theoretical model based on local generation of density fluctuations via nonlinear interactions of Alfven waves propagating away from the Sun.

\begin{center}
{\large\bf
Generation of MHD turbulence non-equilibrium ion distributions
}

{\it
H. J. Fahr
}

Argelander Institut f\"ur Astronomie, University of Bonn, Germany
\end{center}

The heliospheric interface plasma, the so-called inner and outer heliosheath, is characterized by specific MHD turbulence levels which for example are highly relevant for the modulation of galactic cosmic rays outside of the solar wind termination shock. As we shall show, these turbulences not only are due to those turbulences convected over the inner SW termination and the outer bowshock into the sheath region, but also due to the generation of turbulence power by non-linear wave particle interactions. We shall show that ion distributions arise in the heliosheath that are unstable with respect to driving MHD waves with their free energies. We shall give estimates on the expected turbulence levels.

\begin{center}
{\large\bf
The Energy Balance in the Solar Wind Formation Region
}

{\it
S. I. Molodykh
}

Institute of Solar-Terrestrial Physics SB RAS, Russia
\end{center}

The nonthermal broadening of coronal lines observedï within short distances from the Sun was analyzed. The analysis revealed that the nonthermal broadening of coronal lines near the Sun is caused by Alfven waves.

Within the MHD approximation, the wave energy flux required for solar wind formation, and also the plasma velocity and temperature were calculated. Electron density distributions and the flow geometry were used as input data. It is shown that the energy flux required for solar wind formation enters the solar corona in the form of Alfven waves whose dissipation provides the heating of the solar wind plasma near the Sun. Their transformation to acoustic waves in this region is less effective than the dissipation. With distance from the Sun the dissipation of the Alfven waves falls off, and the heating of the solar wind plasma is determined by the coefficient of transformation of the Alfven waves to acoustic waves. Subsequently, the dissipation effectiveness of the acoustic waves decreases, and as soon as the absorption coefficient of acoustic waves becomes less than the transformation coefficient of the Alfven waves to acoustic waves, plasma heating is now determined by the absorption of acoustic waves.

\begin{center}
{\large\bf
Coronal scattering of radio emission under strong regular refraction
}

{\it
A. Afanasiev
}

Institute of Solar-Terrestrial Physics, Russia
\end{center}

It is customary to investigate the scattering of radio emission
from external sources, when probing the near-solar plasma, under
the assumption of a spherically symmetric model for the regular
profile of electron density in the corona. With this assumption,
in the case of large elongations of the source the influence of
regular refraction on the scattering leads only to quantitative
changes in fluctuation characteristics of radio emission. However,
when probing the corona in the case of small elongations, a
regular caustic can be formed, which introduces significant
changes into the scattered field structure. It is important to
analyze statistical characteristics of radio emission that is
scattered in the corona, in the neighborhood of a regular caustic,
because in this case the destruction effect of a regular caustic
under the influence of the corona's turbulent inhomogeneities is
useful for their diagnostics. Strong regular refraction has a
substantial influence on statistical characteristics of radio
emission from intrinsic coronal sources. As is known, in the
neighborhood of such sources there can exist different large-scale
regular electron density features (coronal arches, streamers,
etc.), giving rise to regular caustics and multipathing of radio
emission. The appearing refraction effects should be taken into
account in the analysis of generation mechanisms of radio emission
from coronal sources.

In this paper, based on an integral representation of the wave field as an interference integral, we have obtained asymptotic expressions for statistical moments of monochromatic and pulsed radio emission that are applicable for the formation conditions of regular caustics and multipathing. Numerical simulation results are presented for statistical characteristics of scattered radio emission from non-solar and coronal sources. Computing results are compared with experimental data.

\begin{center}
{\large\bf
Subdiffusion of beams through interplanetary and interstellar media
}

{\it
A. A. Stanislavsky
}

Institute of Radio Astronomy, Ukraine
\end{center}

The angular distribution of beams, being propagated through a medium with random inhomogeneities, is analyzed. The peculiarity of this medium is that beams are trapped at random locations and random times because of wave localization in the inhomogeneities. The equation for the angular distribution is derived. The mean square deviation of the beam from its initial direction is calculated. The application of this method to the diagnostics of interplanetary and interstellar turbulent media is discussed.

\begin{center}
{\large\bf
Investigations of AGNs by the interplanetary scintillation method
}

{\it
V. S. Artyukh
}

Pushchino Radio Astronomy Observatory, Lebedev Physical Institute, Russia
\end{center}

This is a review of investigations of active galactic nuclei (AGNs), that contain compact radio sources, made in PRAO. The investigations are based on the low frequency observations of compact radio sources in the AGNs. The observations were made at a frequency 102 MHz with the Large Phased Array by the interplanetary scintillation method.

\begin{center}
{\large\bf
Investigation of different types of radio sources by IPS method
}

{\it
S. A. Tyul'bashev
}

Pushchino Radio Astronomy Observatory, Lebedev Physical Institute, Russia
\end{center}

Interplanetary scintillation observations of compact steep spectrum sources, flat spectrum sources, gigagherts peaked spectrum sources and core dominated sources were carried out at 111 MHz using interplanetary scintillation method on the Large Phased Array (LPA) in Pushchino. We were able to estimate flux densities or upper limits of flux densities of more than 200 sources.
The physical conditions (magnetic field, density of relativistic particles, energies of magnetic field and relativistic particles) in compact details of investigated sources were estimated. Test of hypothesis of equipartion between energy of magnetic field and energy relativistic particles was made.

\begin{center}
{\large\bf
The dual-frequency calibration of ionosphere influence in VLBA data processing
}

{\it
A. A. Chuprikov
}

Astro Space Center of P.N. Lebedev Physical Institute, Russia
\end{center}

We have used the combined S/X VLBA observation data kindly placed into our deposition by Dr. Leonid Petrov for creation and testing of the dual-frequency calibration procedure has been implemented into the project "Astro Space Locator". Description of this procedure and results of its usage in VLBA data processing for the solving of ionosphere problem are presented.

\begin{center}
{\large\bf
Monitoring of interplanetary and ionosphere scintillations
at frequency 110 MHz
}

{\it
V. I. Shishov, S. A. Tyul'bashev, I. A. Subaev
}

Pushchino Radio Astronomy Observatory, Lebedev Physical Institute, Russia
\end{center}

Large Phased Array of P.N. Lebedev Physical Institute  has the highest in the world  sensivity at meter waveband. It allows to observe simultaneously in two independent diagrams with 16 beams each. We plan to organize a monitoring of interplanetary  and ionospheric scintillation using observations of radio sources by one multi beam diagram. The methods observations and reductions are discussed.  First results of test observations during 7 days of a strip of the sky in a region with declinations between 28o and  32o  with using of  8 beams  are presented.

{\it Acknowledgements}
Authors are grateful for a support of the work by  and grant No 04-02-17332 of Russian Foundation for Basic Research.

\end{document}